\documentclass[conference]{IEEEtran}
\IEEEoverridecommandlockouts

\usepackage{cite}
\usepackage{amsmath,amssymb,amsfonts}
\usepackage[procnumbered,linesnumbered,ruled]{algorithm2e}
\usepackage{algorithmic}
\usepackage{graphicx}
\usepackage{textcomp}
\usepackage{xcolor}
\usepackage{pifont}
\usepackage{wrapfig}
\usepackage{comment}
\usepackage{siunitx}
\usepackage{hyperref}

\def\BibTeX{{\rm B\kern-.05em{\sc i\kern-.025em b}\kern-.08em
    T\kern-.1667em\lower.7ex\hbox{E}\kern-.125emX}}

\newcommand{\eat}[1]{{}}
\newcommand{\para}[1]{\smallskip \noindent\textbf{#1}}
\newcommand{\softpara}[1]{\smallskip \noindent \underline{#1}}

\newcommand{\acp}{\texttt{ACP}}
\newcommand{\odo}{\texttt{ODO}}
\newcommand{\ucp}{\texttt{UCP}}

\newcommand{\mpfont}{\scriptsize}

\ifx\noeditingmarks\undefined
    \newcommand{\MPworker}[2]{{\color{#1}\vrule\vrule}{\marginpar{\color{#1}\mpfont #2}}}
\else
    \newcommand{\MPworker}[2]{}
\fi

\newcommand{\noeditingmarks}{}

\ifx\noeditingmarks\undefined
    
\else
    
\fi

\ifx\noeditingmarks\undefined
    \newcommand{\red}[1]{\textcolor{red}{#1}}
\else
    \newcommand{\red}[1]{\textcolor{black}{#1}}
\fi

\begin{document}
\bstctlcite{IEEEexample:BSTcontrol}

\title{Design and Simulation of the Adaptive Continuous Entanglement Generation Protocol}

\author{Anonymous Authors}

\author{
\IEEEauthorblockN{Caitao Zhan$^{\dagger}$, Joaquin Chung$^{\dagger}$, Allen Zang$^{*}$, Alexander Kolar$^{*}$, Rajkumar Kettimuthu$^{\dagger}$}

\IEEEauthorblockN{$^{\dagger}$Argonne National Laboratory (USA), $^{*}$The University of Chicago (USA)}

\thanks{This material is based upon work supported by the U.S. Department of Energy, Office Science, Advanced Scientific Computing Research (ASCR) program under contract number DE-AC02-06CH11357 as part of the InterQnet quantum networking project. 
}   
}

\maketitle

\begin{abstract}
Generating and distributing remote entangled pairs (EPs) is a primary function of quantum networks, 
as entanglement is the fundamental resource for key quantum network applications. 
A critical performance metric for quantum networks is the time-to-serve (TTS) for users' EP requests, which is the time to distribute EPs between the requested nodes. 
Minimizing the TTS is essential given the limited qubit coherence time. 
In this paper, we study the Adaptive Continuous entanglement generation Protocol (ACP), which enables quantum network nodes to continuously generate EPs with their neighbors, while adaptively selecting the neighbors to optimize TTS. 
Meanwhile, entanglement purification is used to mitigate decoherence in pre-generated EPs prior to the arrival of user requests.
We extend the SeQUeNCe simulator to fully implement ACP and conduct extensive simulations across various network scales. 
Our results show that ACP reduces TTS by up to $94\%$ and increases entanglement fidelity by up to $0.05$.

\end{abstract}

\begin{IEEEkeywords}
Quantum Network, Simulator, Continuous Entanglement Generation, Entanglement Distribution, Protocol
\end{IEEEkeywords}
\section{Introduction}
\label{sec:intro}

Quantum communications hold enormous potential for many ground-breaking scientific and technological advances. 
A quantum network~\cite{wehner2018quantum} serves as the backbone for applications such as 
distributed quantum computing~\cite{calaffi-dqc-cn24} and distributed quantum sensing~\cite{zang-dqs,zhan-opt-24,hillery-qsn-pra23,zhan-localization-qce23}. 
A core functionality of a quantum network is to distribute entangled pairs (EPs) between two distant network nodes, providing a key resource for quantum network applications.
However, distributing an EP across remote nodes can introduce significant latency due to the probabilistic nature of the underlying physical processes. 
Thus, a key performance metric of a quantum network is the time required to serve a user request.
We call this metric time-to-serve (TTS), which is the time required to successfully distribute end-to-end EPs at or above the requested fidelity.
Minimizing TTS is essential for quantum applications due to the limited qubit lifetime.
This will reduce wait time for quantum network users, while improving overall service quality.

Motivated by the above considerations, the primary goal of this work is to reduce TTS for user requests. 
A common approach to reduce TTS involves optimizing entanglement routing~\cite{caleffi-routing-access17,shi-routing-sigcomm20,ghader-swappingtree-tqe22,abane-routingsurvey}, however, we take a different approach.
We focus on \emph{continuously generating EPs between neighbor nodes} in the background, independent of user requests. 
These pre-generated EPs~\cite{kolar-ac-infocom22,inesta-continuous-pra23,ghader-predistribution-qce22} are then used to expedite the time required to fulfill the requests. 

To further improve the efficiency, we design an \emph{adaptive scheme} that leverages past information 
to dynamically guide nodes 
in selecting more optimal neighbors for EP generation ~\cite{kolar-ac-infocom22}.
We call this approach Adaptive, Continuous entanglement generation Protocol (\acp).
While 
\acp\ improves TTS through pre-generated EPs, the fidelity of these 
EPs may degrade as they remain idle in quantum memories before being used to serve future requests.
To mitigate this, we use entanglement purification~\cite{bbpssw} to improve the fidelity of the pre-generated EPs. 

We make substantial extensions to the discrete-event quantum network simulator, SeQUeNCe~\cite{sequence}, to support the implementation of the \acp.
The three key extensions are: 1) a Single Heralded Entanglement Generation Protocol that is aware of pre-generated EPs, 2) a Resource Reservation Protocol to make reservations for the \acp, and a Resource Manager that configures the entanglement purification policy.
We evaluate \acp\ in SeQUeNCe across various network scales by simulating a multi-user quantum network, with user requests sampled from a traffic matrix.
Ou simulation results show that the \acp\ can indeed reduce the request TTS by \red{$57\%\sim94\%$} while simultaneously improving the fidelity by $0.01\sim0.05$ through entanglement purification.

\para{Contributions and Paper Organization.} We define a quantum network model and formulate the problem of adaptive, continuous entanglement generation in~\S\ref{sec:problem}.
Our key contributions are as follows:

\begin{enumerate}
    \item We introduce the Adaptive, Continuous entanglement generation Protocol (\acp) in \S\ref{sec:acp}, which reduces the TTS of user requests. 
    Additionally, entanglement purification (\S\ref{subsec:resource}) enhances the fidelity of EPs.
    \item We extend the SeQUeNCe quantum network simulator (\S\ref{sec:simulator}) with significant enhancements. 
    In particular, a single-heralded entanglement generation protocol is included that accounts for pre-generated EPs. 
    \item We conduct extensive evaluation of \acp\ using SeQUeNCe and demonstrate its effectiveness in reducing TTS and improving the fidelity of distributed EPs in \S\ref{sec:result}.
\end{enumerate}

\section{Model, Problem, and Related Work}
\label{sec:problem}

In this section, we define the quantum network model and the problem, as well as discuss the related work.

\subsection{Quantum Network Model} 
We define a quantum network as a graph $G=(V, E)$, with $V=\{v_1, ..., v_n\}$ and $E=\{e(v_i, v_j)\}$ denoting the set of nodes and links, respectively.
Pairs of nodes connected by a link are defined as \emph{neighbor} nodes.
An EP between neighbors (i.e., on a link $e(u, v)$) is called an ``elementary link EP'' or a ``link EP'' for short.
Each node has a finite number of emissive quantum memories.
These memories generate atom-photon entanglement, where the atomic part of the EP is stored in the memory and the photonic part is transmitted to a distant device for photonic Bell-state measurement (BSM).
The atomic part of the EP has a time-dependent decoherence model.
Moreover, each memory has a maximum photon emission rate and multiple memories can emit photons simultaneously.
Entanglement swapping and purification are performed on the atomic memory qubits.

Each link $e(u, v)$ in the quantum network is comprised of two quantum channels and four classical channels.
The two quantum channels connect nodes $u$ and $v$ to the middle photonic BSM device, where photons emitted by quantum memories are measured.
Two of the four classical channels connect the photonic BSM device to both $u$ and $v$ to send the measurement results,
while the other two classical channels connect $u$ to $v$ and vice versa to send messages related to protocol coordination.
We assume lossless classical communication, while quantum bits have three sources of loss (see Table~\ref{tab:sim-parameter} in \S\ref{sec:result}): (1) failure during the memory emission, (2) loss during the transmission in the quantum channel, and (3) failure of the detector in the photonic BSM device. 

\subsubsection{User Request and Time-to-serve (TTS)} 
Fig.~\ref{fig:request} shows an example of a \emph{user request} arriving
at an \texttt{initiator} node, with the following attributes: \texttt{responder}, \texttt{start time}, \texttt{end time}, \texttt{number of EPs}, and \texttt{fidelity}.
The request asks the quantum network to generate a certain number of end-to-end EPs between the initiator node and the responder node with fidelity above a threshold within a time range [\texttt{start time}, \texttt{end time}].
If the generation of the requested end-to-end EPs succeeds at time $t$ and $t$ is within the time range, then the request's time-to-serve (TTS) is
\begin{equation}
    \mathrm{TTS} = t - \mathrm{start~time}
    \label{eqn:tts}
\end{equation}
i.e., the time elapsed from the request start time to the time $t$ when all the end-to-end EPs are distributed.
The TTS includes the time for elementary link entanglement generation and entanglement swapping.

\subsubsection{Entanglement Path Computation} 
\begin{figure}
    \centering
    \includegraphics[width=0.75\linewidth]{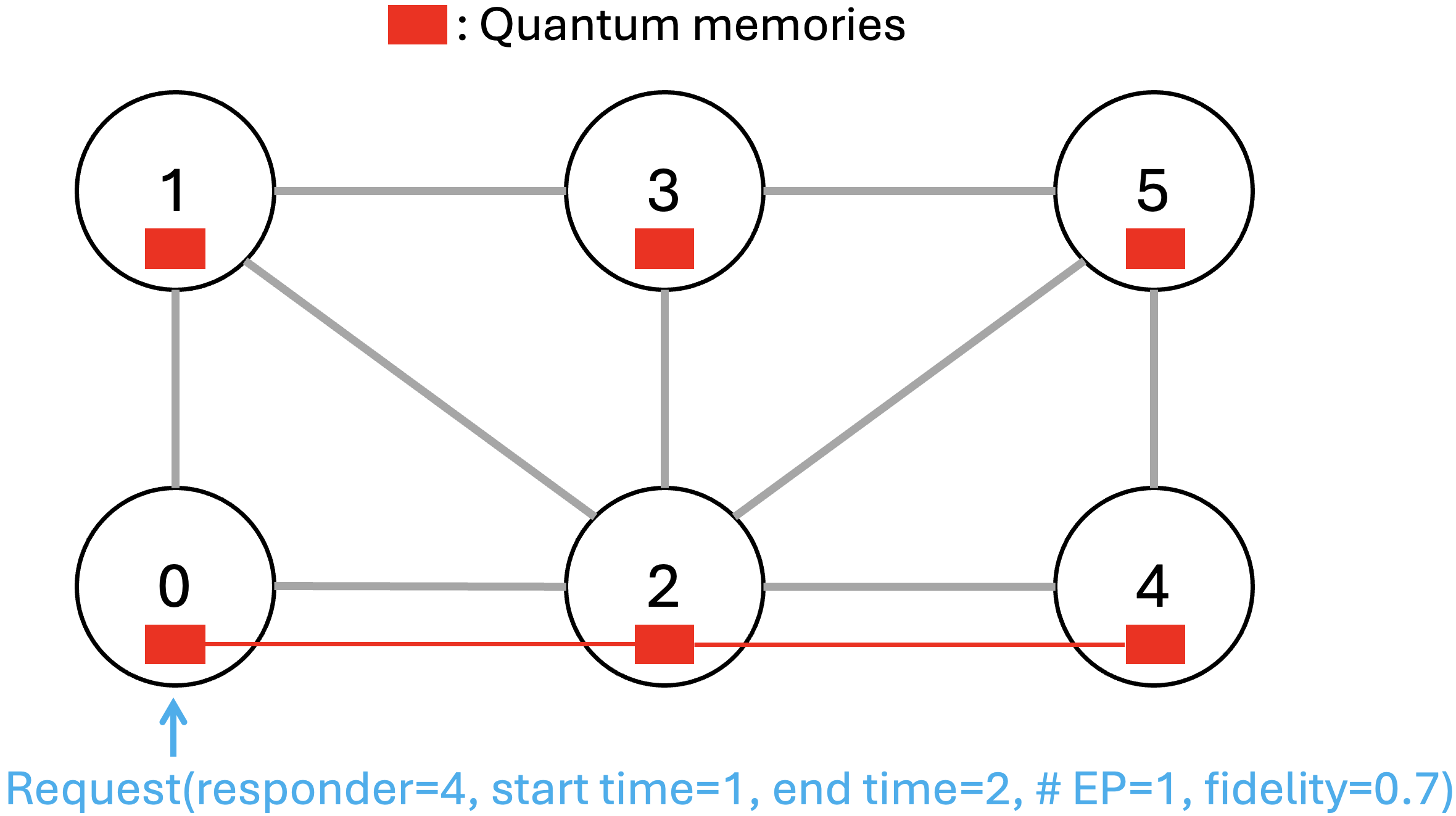}
    \vspace{-0.1in}
    \caption{Toy example of a quantum network, request, and time to serve (TTS).
    \red{A request arrives at node 0 (initiator), asking node 0 to generate 1 EP with node 4 (responder) within the [1, 2] seconds time range, and requires the fidelity greater than 0.7.}
    If the EP is generated at time $t = 1.2$ seconds, then the TTS of this request is 0.2 seconds.}
    \vspace{-0.1in}
    \label{fig:request}
\end{figure}

The path computed by the entanglement routing algorithm directly affects the TTS. 
For example, consider end nodes nodes 0 and 4 in Fig.~\ref{fig:request}; 
the path (0,2,4) will have a smaller expected TTS than path (0,1,2,5,4).
In this work, 
\red{we assume all nodes use static routing (i.e., the forwarding table of each node is pre-configured following the shortest distance path and remains fixed throughout the simulation).}
We assume there is a small time interval between the request arrival at the initiator node and the request's start time.
The path computation starts after the request's arrival and is quickly completed before the request's start time.
Thus, the time for path computation is not included in the TTS. 

\subsubsection{Motivation} 
We observe that a major component of the TTS is the time to generate link EPs. 
We aim to reduce this time by pre-generating EPs before the request's start time,
such that when a request arrives, pre-generated EPs can be immediately used.
However, nodes do not have the request's information beforehand; thus, nodes ought to continuously generate EPs with their neighbors regardless of incoming requests.
Inspired by~\cite{kolar-ac-infocom22}, an adaptive control scheme is introduced to guide the nodes to select more frequently used links to improve TTS.

\subsection{Problem Statement} 
\begin{figure}
    \centering
    \includegraphics[width=0.9\linewidth]{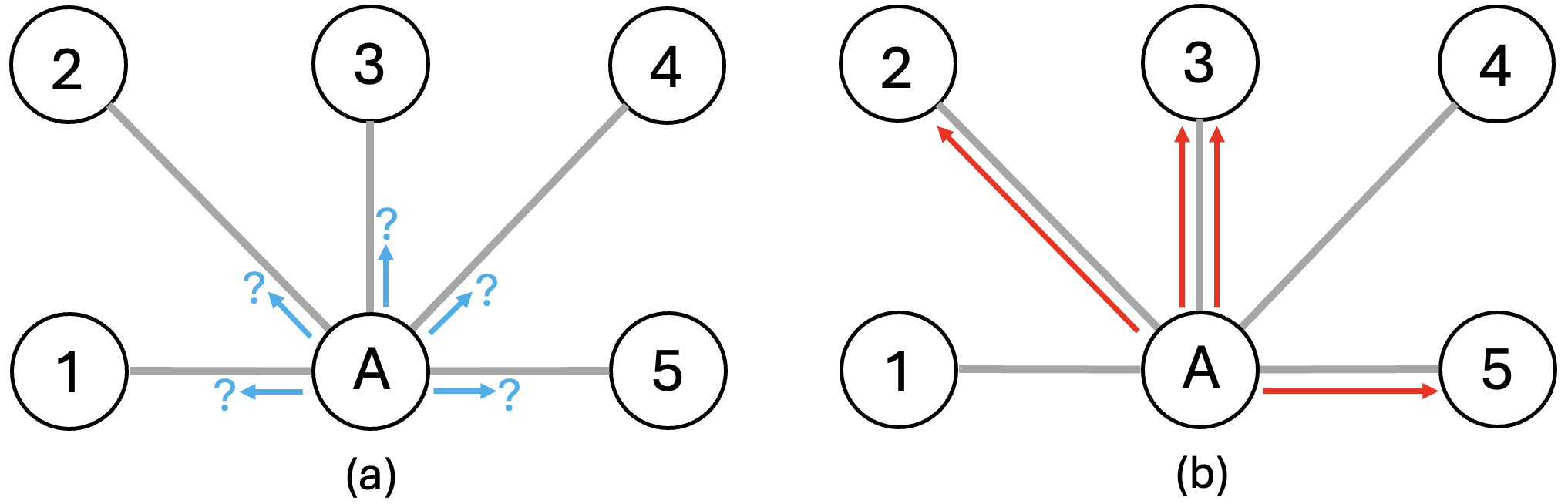}
    \vspace{-0.1in}
    \caption{Toy example of the problem. The \acp\ runs on Node-A, which has five neighbors in total. \acp\ is allowed to use a maximum of four quantum memories for continuous generation of link EPs. Node-A in (a) has to decide what neighbors to generate EP with.
    In (b), the user requests result in paths (computed from the entanglement routing algorithm) that frequently include the segments [2, A, 3] and [3, A, 5]. Thus, Node-A should select neighbor Node-3 the most, followed by Node-2 and Node-5.}
    \vspace{-0.1in}
    \label{fig:acp-highlevel}
\end{figure}

We assume a quantum network with a \emph{distributed}, \emph{online}, and \emph{asynchronous} setting:
\begin{itemize}
    \item \textit{Distributed:} Each node only has a local view of the network (i.e., its neighbors).
    \item \textit{Online:} Each node can be adaptive to request arrival patterns and available elementary link EPs, which are unknown in advance.
    \item \textit{Asynchronous:} 
    The design and simulation of protocols operate in discrete events (i.e., no time slots are assumed).
\end{itemize}
Under the above setting, the problem is to design and implement a continuous entanglement generation protocol, working in conjunction with an entanglement routing protocol, to reduce the request's TTS.
The proposed protocol will allow a node to continuously generate EPs with its neighbor regardless of request arrivals.
As a result, when a request arrives and a path has been computed, pre-generated EPs could be immediately used instead of generating EPs on demand after the request's start time.
A naive protocol could randomly select neighbors to generate EPs with.
However, an adaptive control that leverages the request information from the past can guide a node to select neighbors more wisely.
This approach will generate link EPs that have a higher chance of being used to serve future requests, thus improving the efficiency of continuous EP generation (see Fig.~\ref{fig:acp-highlevel}).

\subsection{Related Work}

The \acp\ is complementary to entanglement routing, which can be performed in an on-demand or continuous manner. 

In on-demand entanglement routing, entanglement generation only starts after a request arrives~\cite{chakra-routing-19}. The typical objective is to find an optimal path, together with a swapping policy (i.e., the order of entanglement swapping), to satisfy throughput, latency, and fidelity requirements for a user request~\cite{abane-routingsurvey}.
Caleffi~\cite{caleffi-routing-access17} proposed an exponential runtime algorithm that examines all possible paths with a metric assuming a balanced tree swapping policy. Shi and Qian~\cite{shi-routing-sigcomm20} created a Dijkstra-like algorithm with a metric considering multiple quantum channels between neighbor nodes and they considered a sequential swapping policy. Ghaderibaneh~\cite{ghader-swappingtree-tqe22,fan-purification} used dynamic programming to find the optimal ``combination of path and swapping policy'', i.e., ``swapping tree''. 

On-demand entanglement routing has also been explored beyond the bipartite regime. In general, the distribution of multipartite entanglement across a network~\cite{pirker-multipartite-19,fan-graphstate} starts from the generation of small entangled states shared by few parties (e.g., normal 2-qubit Bell state or 3-qubit GHZ states~\cite{lee-graphstate}). Then multipartite entanglement routing mainly deals with the optimal policy for assembling the small entangled states into large multipartite entangled states.

In continuous entanglement routing, elementary link EPs are generated continuously over all links in the background~\cite{abane-routingsurvey}. Then, the routing algorithm computes the path on the instant logical topology formed by the created link entanglements~\cite{cicco-routing-21}. Kolar et al~\cite{kolar-ac-infocom22} first considered continuous generation of elementary link EPs guided by an adaptive scheme, but the evaluation was ad hoc and quantum memory decoherence was not considered. 

Continuous entanglement generation can also be extended from between neighbor nodes (physical link) to between non-neighbor nodes (virtual link), with the help of entanglement swapping. This allows the network to go beyond the physical connection and create arbitrary virtual topologies, which offers the potential to increase the connectivity of the network, reduce the diameter of the network, and ultimately reduce the latency of end-to-end entanglement generation~\cite{ghader-predistribution-qce22}. Inesta and Wehner~\cite{inesta-continuous-pra23} defined two metrics to measure the quality of continuous entanglement generation and distribution protocols: (1) virtual neighborhood size and (2) virtual node degree. The higher the values of the two metrics are, the better the continuous protocol is.

\begin{figure*}[ht]
    \centering
    \includegraphics[width=0.96\linewidth]{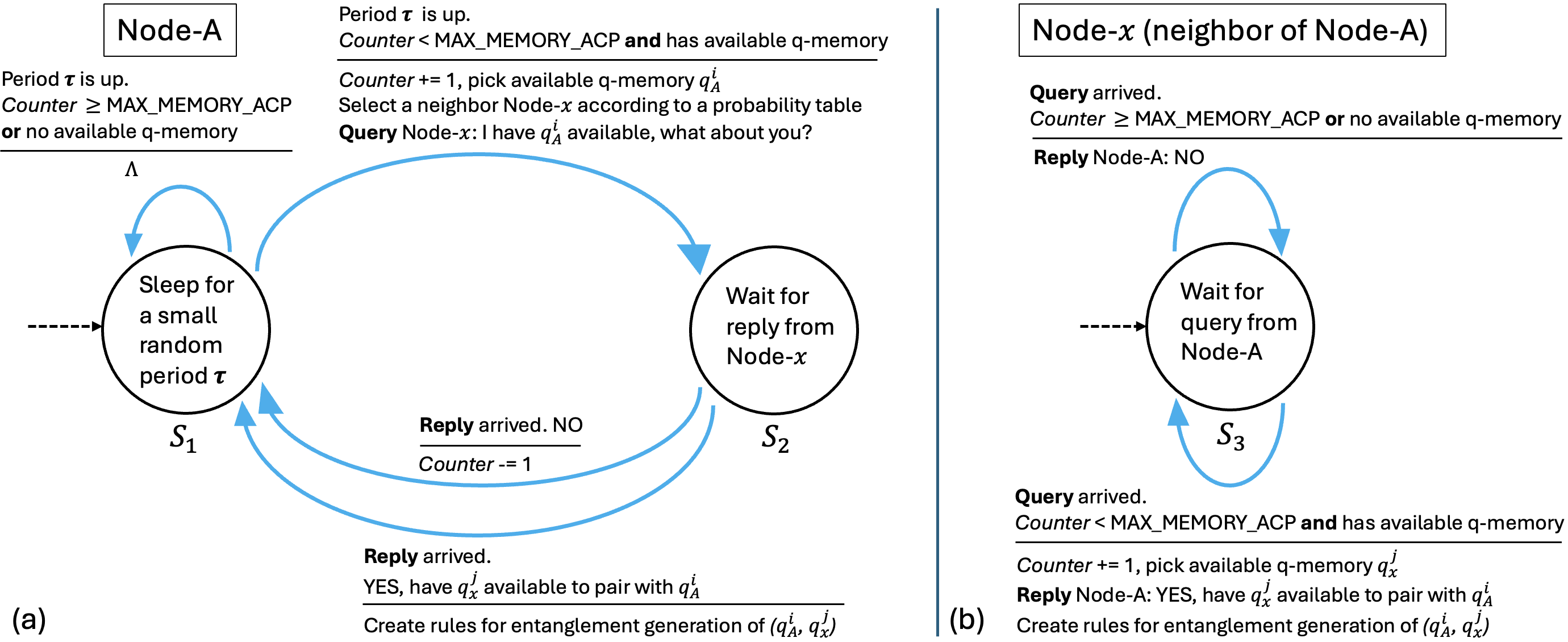}
    \vspace{-0.1in}
    \caption{The two FSMs of \acp. (a) depicts the FSM at Node-A, while (b) shows the FSM at Node-A's neighbors.
    Note that \acp\ is a symmetric \red{peer-to-peer} protocol, so the FSM in (a) is also running on Node-A's neighbor, and the FSM in (b) is also running on Node-A.
    \red{This figures depicts ``one connection''. We achieved multiple connections by having multiple $S_2$ states in the Node-A FSM.}
    }
    \vspace{-0.1in}
    \label{fig:acp-fsm}
\end{figure*}

\section{Adaptive, Continuous Entanglement Generation Protocol}
\label{sec:acp}

In this section, we formulate the \acp, which allows a node to continuously generate EPs with their neighbors, with neighbor selection guided by an adaptive control.
The \acp\ runs in the background in complement to entanglement routing protocols for on-demand requests.

\subsection{Continuousness and Finite State Machine}
\label{subsec:continuousness}

We use finite state machines (FSMs) to describe the continuousness of the \acp.
A FSM is defined by \emph{state}, \emph{transition}, \emph{event}, and \emph{action}.
A state describes the system's present situation, including what it is doing or waiting for. 
It is depicted as a circle, and the circle's text describes the behaviors or processes that occur while the system is in the state (see Fig.~\ref{fig:acp-fsm}).
A transition is the change from one state to another and is depicted by an arrow.
An event causing the transition is shown above the horizontal line labeling the transition, and the actions taken when the event occurs are shown below the horizontal line~\cite{topdown-approch}.
We use two FSMs to define the \acp: one for Node-A and one for Node-A's neighbor Node-x.
\red{It resembles the classic ``sender-receiver'' FSMs~\cite{topdown-approch} where Node-A is the sender and Node-x is the receiver.}

\para{States.}
\red{The two states of the FSM at Node-A are}:
\begin{itemize}
    \item $S_1$: Sleep for a small random period $\tau$.
    \item $S_2$: Wait for a reply from Node-$x$.
\end{itemize}
\red{The state of FSM at Node-$x$ (neighbor of Node-A) is:}
\begin{itemize}
    \item $S_3$: Wait for a query from Node-A.
\end{itemize}

\para{Transition.} Three factors determine the transition:
\begin{enumerate}
    \item \emph{Counter}: each node has a counter to keep track of the number of quantum memories occupied by the \acp.
    \item \emph{MAX\_MEMORY\_ACP}: determines the maximum number of quantum memories  that the \acp\ is allowed to occupy.
    This number must be smaller than the total number of quantum memories at the node.
    \item Available quantum memory $q_A^i$ ($i$th memory at node-A) not occupied by any protocols, including \acp\ as well as entanglement generation protocol, entanglement swapping protocol, and purification protocol (see Section~\ref{sec:simulator}). 
\end{enumerate}

\softpara{$S_1 \rightarrow S_2$ and $S_1 \rightarrow S_1$}: At Node-A, the FSM starts at state $S_1$, where the FSM sleeps for a small random period $\tau$.
When the period $\tau$ is up, the FSM will take different actions depending on the condition.
If the condition ``\emph{Counter} is less than \emph{MAX\_MEMORY\_ACP} and Node-A has available q-memories'' is satisfied, then the FSM will take several actions and it will transition to state $S_2$, where it waits for a reply from Node-$x$.
The actions include: incrementing \emph{Counter} by one, picking an available q-memory (e.g., $q_A^i$), selecting a neighbor Node-$x$ according to a probability table (see Fig.~\ref{fig:acp-probtable}(a)), and querying the availability of a q-memory to Node-$x$ while notifying it that $q_A^i$ has been picked on Node-A.
If the condition is not satisfied, the FSM will do nothing and transition back to $S_1$.

\softpara{$S_3 \rightarrow S_3$}: At Node-$x$, the FSM starts at state $S_3$, where it waits for a query from Node-A. 
When a query arrives, the FSM will transition back to $S_3$, and it performs different actions depending on the condition. 
If the condition ``\emph{Counter} is smaller than \emph{MAX\_MEMORY\_ACP} and Node-$x$ has available q-memory'' is satisfied, the FSM will increment \emph{Counter} by one, pick an available q-memory (e.g., $q_x^j$), reply to Node-A stating \emph{YES} and that $q_x^j$ is available to pair with $q_A^i$, and create an entanglement generation protocol for $(q_A^i, q_x^j)$.
After the entanglement generation protocol finishes in the future, \emph{Counter} is decremented by one (not shown in the FSM in Fig.~\ref{fig:acp-fsm}).
If the condition is not satisfied, the FSM replies with a message stating \emph{NO}.

\softpara{$S_2 \rightarrow S_1$}: When Node-$x$'s reply arrives at Node-A, the FSM will transition to $S_1$ and perform different actions depending on the reply's message. 
If the reply is \emph{YES} with $q_x^j$ being available to pair with $q_A^i$, \red{Node-A will create a rule that generates entanglement generation protocol for $(q_A^i, q_x^j)$.
After the rule expires, the memory is released and \emph{Counter} is decremented by one.}
If the reply is \emph{NO}, then \emph{Counter} is decremented by one.

\subsection{Adaptive Control}
\label{subsec:adaptive-control}
\begin{figure}[t]
    \centering
    \includegraphics[width=\linewidth]{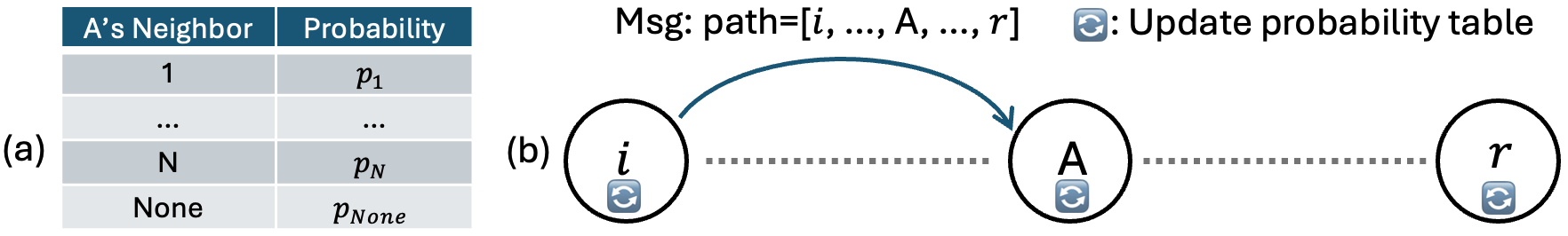}
    \vspace{-0.2in}
    \caption{(a) Node-$A$'s probability table. (b) Node-$i$ receives a request to generate EPs with Node-$r$. 
    After the request is served, Node-$i$ and Node-$r$ will update their probability tables. 
    Meanwhile, Node-$i$ sends a message containing the path to the intermediate nodes.
    Intermediate nodes update their probability tables after receiving the message. }
    \vspace{-0.1in}
    \label{fig:acp-probtable}
\end{figure}

The adaptive control of \acp\ at each node dynamically updates a local \emph{probability table} in real-time to adjust to request patterns. 
When a node needs to select a neighbor for pre-generating entanglement, it runs a roulette wheel selection algorithm informed by its probability table.
See Fig.~\ref{fig:acp-probtable}(a) for an example of a probability table.
The first column of the probability table is the neighbor, and the second column is the probability of selecting that neighbor.
Our implementation requires a ``phantom'' neighbor denoted \emph{None} that is also associated with a probability.
In an online setting, the probability table can adapt to user request patterns, which are unknown beforehand.

\para{Adaptiveness Mechanism.} We want Node-A to have a higher chance of selecting neighbor Node-$x$ if the link (Node-A, Node-$x$) frequently appears in the entanglement path of the user requests.
The adaptive control will update the probability tables after each request is served.
When Node-$i$ (request initiator) receives a request to generate an EP with Node-$r$ (request responder), the entanglement routing protocol will compute a path and start the on-demand entanglement generation and swapping along the path.
The request is served after the end-to-end EP is distributed.
Once the request is served, Node-$i$ and Node-$r$ will use Algorithm~\ref{algo:adaptive} to update their local probability tables directly;
as end nodes, Node-$i$ and Node-$r$ have the path information $p=[i, \cdots, r]$.
After Node-$i$ updates its probability table, Node-$i$ will send a message to all intermediate nodes containing $p$ (see Fig.~\ref{fig:acp-probtable}(b)).
Upon receiving the message containing $p$, each intermediate node will run Algorithm~\ref{algo:adaptive} to update their probability table.

\para{Algorithm~\ref{algo:adaptive}} is designed to give a reward to $P_x$ of Node-A's probability table by adding a small positive number $\delta$ to $P_x$ if the link (Node-A, Node-$x$) is indeed part of the entanglement path of the request.
Thus, the \acp\ will have a higher chance of generating EP for the link (Node-A, Node-$x$), assuming future requests follow a similar path.
\red{A larger value of $\alpha$ lets the algorithm adapt to the new observations (new EPs) more quickly, while also making the algorithm less stable. We observe that a value around 0.05 is a good balance.}

\begin{algorithm}[t] 
        \KwIn{This node $Node\text{-}A$, entanglement path $path$}
        \KwIn{Probability table $P$, delta $\delta$}
        \KwOut{$P$, updated probability table of this node}
        $neighbors \leftarrow$ All neighbors of $Node\text{-}A$ \\
	\For{$Node$-$x$ in neighbors}{
           \If{Link ($Node\text{-}x$, $Node\text{-}A$) in $path$}{
                $P_x \leftarrow P_x + \delta$ 
           }
        }
        $P \leftarrow$ Normalized $P$ \ \ \ \ \tcp{$\sum P_* = 1$} 
        \Return $P$
	\caption{Update($Node\text{-}A$, $path$, $P$, $\delta$)}
\label{algo:adaptive}
\end{algorithm}



\section{Extensions of the SeQUeNCe Simulator}
\label{sec:simulator}

Recently, several quantum network simulators have been developed such as SeQUeNCe~\cite{sequence}, QuISP~\cite{quisp}, and NetSquid~\cite{netsquid}.
We chose SeQUeNCe for our simulation because it is open-source, it is easy to understand (Python source code), it has good documentation, and most importantly it is customizable and easy to extend.
In this section, we present the extensions of SeQUeNCe we make to support the implementation of the \acp.
The extensions are open-source at {\url{https://github.com/caitaozhan/adaptive-continuous}}.

\subsection{SeQUeNCe Overview}
\begin{figure}[b]
    \vspace{-0.2in}
    \centering
    \includegraphics[width=0.75\linewidth]{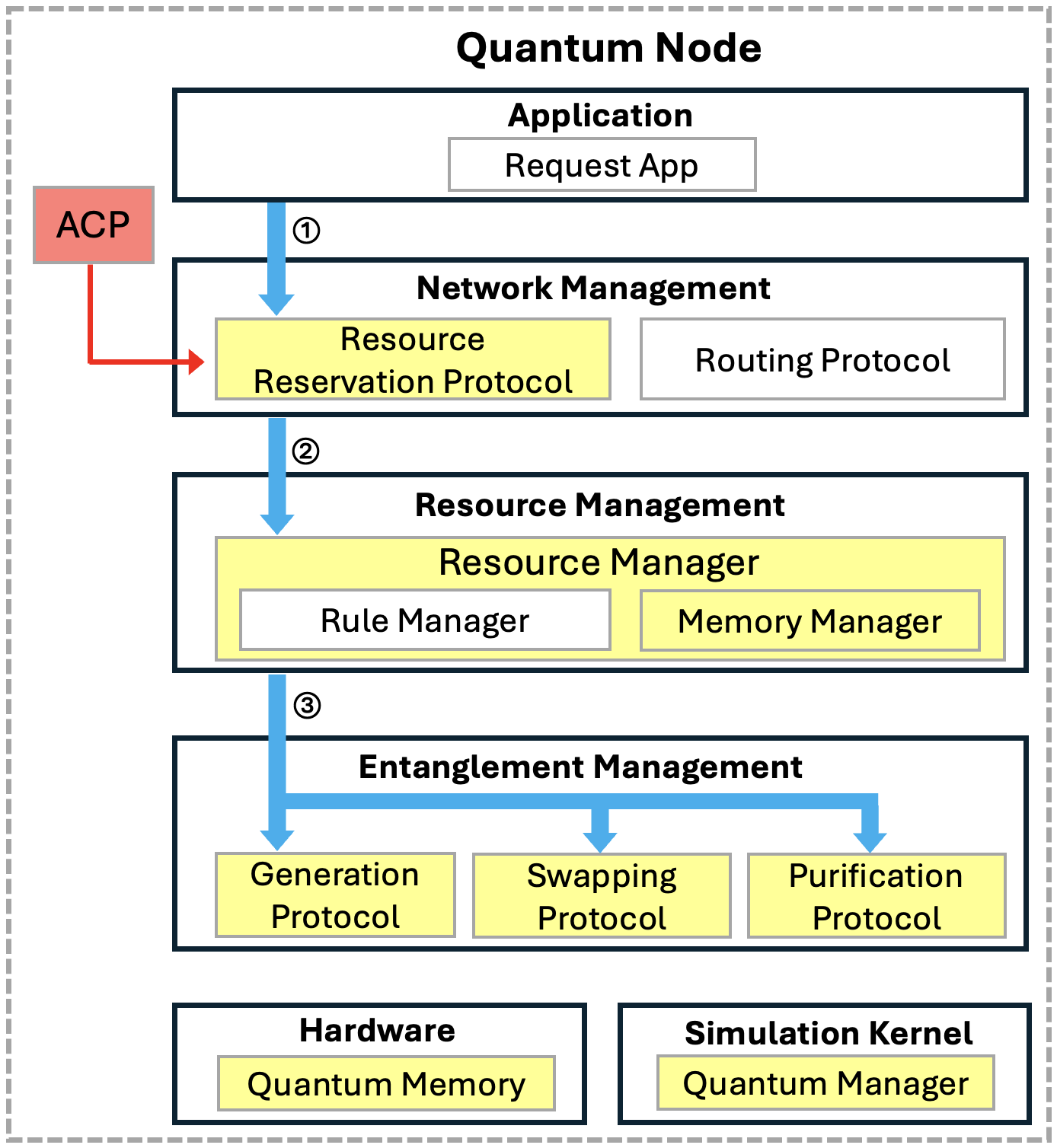}
    \vspace{-0.1in}
    \caption{SeQUeNCe architecture has six modules.
    Each module has several components (not all components are shown).
    Subsection~\ref{subsec:quantum_manager} to~\ref{subsec:reservation} discuss the extensions in the eight highlighted components in yellow.}
    \vspace{-0.1in}
    \label{fig:sequence}
\end{figure}

SeQUeNCe adopts a modular design~\cite{sequence} that separates functionality into modules to support the development of future quantum network architectures, new hardware~\cite{zang2022simulation,davossa2024simulation}, new protocols, and new applications while allowing parallelization~\cite{wu2024parallel}.
Fig.~\ref{fig:sequence} shows the six main modules of SeQUeNCe: Application, Network Management, Resource Management, Entanglement Management, Hardware, and Simulation Kernel.
Not all components of each module are depicted in Fig.~\ref{fig:sequence}.
For a detailed description of the connections and dependency of the modules and components, we refer the reader to Sections 3 and 4 of~\cite{sequence}.
We highlight three main aspects of SeQUeNCe modules:
\begin{enumerate}
    \item \textit{Application module:}
    It creates requests and sends them to the Network Management module (see \ding{172} in Fig.~\ref{fig:sequence}).
    
    \item \textit{Network Management module:}
    Upon receiving a request, the Resource Reservation protocol creates rules and installs them in the Rule Manager of the Resource Manager (see \ding{173} Fig.~\ref{fig:sequence}).
    
    \item \textit{Resource Management module:}
    When the condition of a rule is satisfied, the rule will create the entanglement protocols (see \ding{174} in Fig.~\ref{fig:sequence}).

\end{enumerate}

\begin{figure*}
    \centering
    \includegraphics[width=0.82\linewidth]{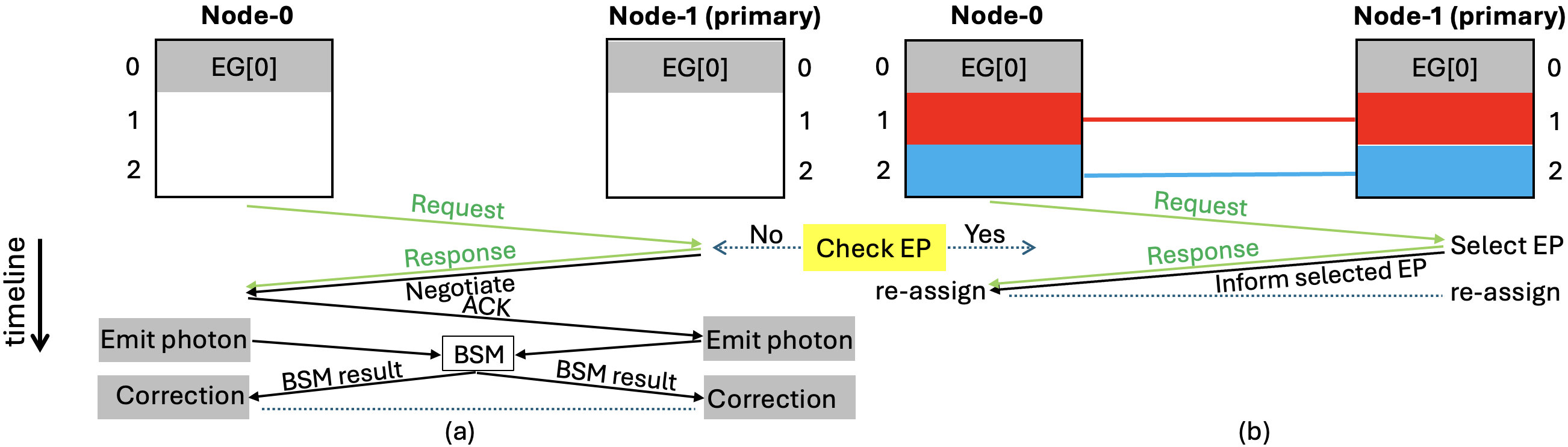}
    \vspace{-0.1in}
    \caption{
    The entanglement generation protocol handles two scenarios (a) and (b).
    The yellow-highlighted ``Check EP'' in the figure determines which scenario will happen.
    (a) shows the scenario without pre-generated EPs (i.e., \acp\ has not generated an EP yet). \red{The photonic BSM device indicates a meet-in-the-middle protocol.}
    (b) shows the case of having pre-generated EPs (red and blue) before a request's arrival.
    Note that the request and response messages in green are the coordination messages of the Resource Manager to connect two entanglement generation protocols at two nodes (e.g., EG[0]), not the user request.}
    \vspace{-0.1in}
    \label{fig:eg}
\end{figure*}

The \texttt{Rule} --which is composed of \emph{priority}, \emph{condition}, and \emph{action}-- plays an important role in SeQUeNCe. 
A Rule will take \emph{action} once the \emph{condition} is satisfied.
If multiple rules are satisfied simultaneously, then the \emph{priority} will break the tie.
Similar to the RuleSet in~\cite{matsuo-ruleset-pra19}, when an action is taken, the entanglement protocols will be created, including generation, swapping, and purification.
Each entanglement protocol is associated with a set of quantum memories on a node.

\subsection{Location of the \acp\ in the SeQUeNCe Architecture}
In Fig.~\ref{fig:sequence}, the \acp\  is placed ``between'' the Application module and Network Management module.
Technically, the \acp\ is implemented outside of all modules and is a direct attribute of a quantum node. 
Logically, the \acp\ belongs to the Application module because the \acp\ calls the Resource Reservation Protocol just like any typical application (e.g., quantum teleportation) by making a request to the Network Management module.
The difference is that typical applications usually make a request on demand due to user behavior, while the \acp\ continuously runs in the background regardless of the user's request.

\subsection{Code Extensions}
This section describes the necessary code extensions for implementing the \acp\ in SeQUeNCe.

\subsubsection{Quantum Manager}
\label{subsec:quantum_manager}
We add a new class \texttt{BellDiagonalState} to the Simulation Kernel to represent a 2-qubit entanglement pair as Bell Diagonal State (BDS).
We also add a new class \texttt{QuantumManagerBellDiagonal} to the Simulation Kernel to track and manage the quantum states with the BDS formalism.
\red{The assumption of BDS is valid as any 2-qubit state can be transformed into one via Pauli twirling with fidelity unchanged.
The value of the first element of the BDS vector equals the fidelity.}

\subsubsection{Quantum Memory}
\label{subsec:memory}
We extend the \texttt{Memory} class with a new method \texttt{bds\_decohere},
which enables time-dependent decoherence for quantum memories according to single-qubit error pattern, i.e., the probability distribution of $X, Y, Z$ Pauli errors $\{p_{X,Y,Z}\}$.
When this method is called, it will first fetch the current time $t_c$ and the time $t_p$ when \texttt{bds\_decohere} was previously called. 
Then the decoherence time $\Delta t$ is computed as $\Delta t=t_c-t_p$.
Given the previous BDS state, $\Delta t$, $\{p_{X,Y,Z}\}$, and the memory coherence time (See Table~\ref{tab:sim-parameter}), the new BDS state after decoherence is analytically computed (see~\cite{zang-dqs} for details). 

\subsubsection{Entanglement Generation Protocol}
\label{subsec:generation}

We add a new class \texttt{EntanglementGenerationACP}.
Compared with SeQUeNCe's original class \texttt{EntanglementGeneration}, the new class is single heralded, it uses BDS to represent the quantum state, and most importantly it is aware of existing EPs pre-generated by the \acp. 
Fig.~\ref{fig:eg}(b) shows two pre-generated EPs in the colors red and blue.
The primary node of the pair (the node that has a larger name alphabetically) will be in charge of selecting one EP among potentially multiple EPs between two neighbor nodes.
Compared with Fig.~\ref{fig:eg}(a) that needs to generate a new EP, Fig.~\ref{fig:eg}(b) shows how reusing existing EPs significantly decreases the latency since link EP generation is probabilistic.
Table~\ref{tab:sim-parameter} shows the parameters used during the simulation.

\subsubsection{Entanglement Swapping Protocol}
We add a new class \texttt{EntanglementSwappingBDS} that uses BDS~\cite{zang-dqs} to represent the quantum states for entanglement swapping and perform analytical derivation of imperfect swapping results.
During entanglement swapping, we assume the Bell state measurement on memory qubits is always successful but will introduce additional noise in gates and measurements. 
See swapping success rate, gate and measure fidelity in Table~\ref{tab:sim-parameter}.

\subsubsection{Entanglement Purification Protocol}
\label{subsec:purification}
We add a new class \texttt{BBPSSWBDS} that uses BDS~\cite{zang-dqs} to represent the quantum states for entanglement purification. 
\red{In \texttt{BBPSSWBDS}, the circuit is fundamentally the same as the BBPSSW protocol~\cite{bbpssw}, while only assuming BDS states instead of Werner states.
}
Entanglement purification has a less-than-one success rate and is analytically calculated given two input EPs.

\subsubsection{Resource Manager}
\label{subsec:resource}
The \texttt{ResourceManager} at each node allows \acp\ to keep track of the EPs it generates with neighbor nodes.
We adopt the common as-soon-as-possible strategy for entanglement purification~\cite{ent_dist_epp}.
Once a new EP between Node-A and Node-$x$ is generated and there exists an older EP between Node-A and Node-$x$ generated by \acp, the \acp\ of the primary node between Node-A and Node-$x$ will be responsible for selecting one of the older EP for purification.
\red{This process is also known as entanglement pumping~\cite{dur2003entanglement,zang2024no}.}
The newer EP will be the kept EP, and the older EP will be the measurement EP.
Given the two EPs, the \texttt{ResourceManager} at the primary node will initiate the creation of the purification protocol. 
First, it creates an entanglement purification protocol locally (the primary node), and then it sends a message containing the information of the two EPs to the non-primary node, so the counterpart protocol can be created therein.

\subsubsection{Memory Manager}
We add the capability of reallocating quantum memories to the \texttt{MemoryManager}.
When a pre-generated EP is chosen to be reused for entanglement generation, the qubits of the pre-generated EP at two nodes must be reallocated.
As shown in Fig.~\ref{fig:eg}(b), when the red pre-generated EP is chosen, it will be reallocated to the gray memory slot, which is allocated for the on-demand EP generation for Application requests.

\subsubsection{Resource Reservation Protocol}
\label{subsec:reservation}
\texttt{create\_rules\_acp} is added to the \texttt{ResourceReservationProtocol} class to create rules only for entanglement generation protocols in response to the \acp.
The Resource Reservation Protocol originally has a method \texttt{create\_rules} that creates the rules for entanglement generation, swapping, and purification in response to the Application request. 
The \acp, however, does not need swapping because it only creates link entanglement pairs with neighbor nodes.
Entanglement purification protocols are created without the rules (see Section~\ref{subsec:resource}).
\section{Simulation}
\label{sec:result}
In this section, we present our simulation results,
which show that \acp\ is very effective in reducing the user request TTS.
Moreover, the fidelity of end-to-end EPs is improved with the help of entanglement purification.

\begin{table}[ht]
    \vspace{-0.1in}
    \caption{Simulation parameter}
    \vspace{-0.1in}
    \centering
    \begin{tabular}{|c|c||c|c|}
        \hline
        End node processing delay & 100 \unit{\us} & Forward delay & 20 \unit{\us}  \\
        \hline
         \# of memory per node & 10  & Fiber attenuation & 0.2 dB/km  \\
        \hline
         Max \# of memory for \acp & 5 & Link distance & 10 km \\
        \hline 
        Quantum memory efficiency & 0.6 &  Speed of light & $2\text{e}8$ m/s \\
        \hline
        Photon detector efficiency & 0.95  & Coherence time & 2 s \\
        \hline
        Initial EP fidelity & 0.95 &Pauli errors & [$\frac{1}{3},\frac{1}{3},\frac{1}{3}$] \\
        \hline
        Swapping success rate & 1 & Gate fidelity & 0.99\\
        \hline
        Photonic BSM success rate & 0.5 & Measure fidelity & 0.99 \\
        \hline
        Request arrival rate & 10 Hz &  $\delta$ in Algo.~\ref{algo:adaptive} & 0.05  \\
        \hline
    \end{tabular}
    \vspace{-0.1in}
    \label{tab:sim-parameter}
\end{table}

\subsection{Simulation Setting}

In this subsection, we discuss the network topology, request pattern, classical communication latency, and the entanglement generation strategies under study.
Table~\ref{tab:sim-parameter} summarizes the key parameters for our simulations. 

\begin{wrapfigure}{r}{1.45in}
    \vspace{-0.2in}
    \hspace{-0.2in}
    \centering
    \includegraphics[width=1.05\linewidth]{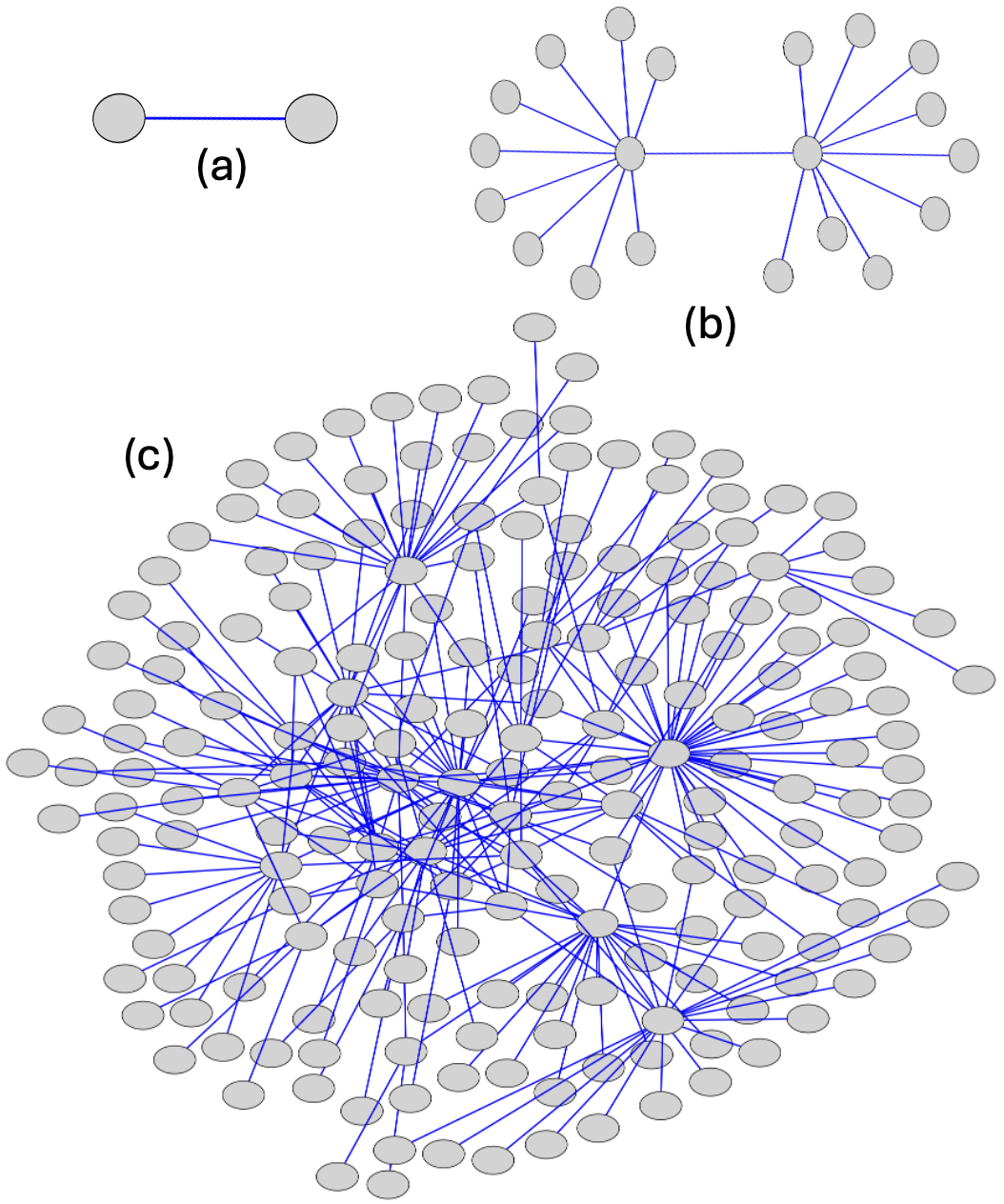}
    \vspace{-0.1in}
    \caption{Network topologies.}
    \vspace{-0.1in}
    \label{fig:topology}
\end{wrapfigure}

\subsubsection{Network Topology} 
We consider three network topologies: a small-scale two-node network, a medium-scale 20-node bottleneck network, and a large-scale 200-node autonomous system (AS) network as shown in Fig.~\ref{fig:topology}(a)$\sim$(c).
We assume that all links have a uniform distance of 10~km, each node has 10 quantum memories, and the \acp\ is allowed to use a maximum of five quantum memories at each node.


\subsubsection{Request Pattern}
Requests are sampled from a \emph{traffic matrix} whose elements sum to one.
The element $[i,j]$ represents the probability of generating a request in which the initiator is Node-$i$ and the responder is Node-$j$.
The request arrival rate is 10~Hz throughout all experiments (i.e., a new request arrives every 0.1 s).
We assume each request asks for one end-to-end EP with fidelity $> 0.5$, and the difference between the \texttt{start time} and \texttt{end time} is slightly $< 0.1$ s (i.e., ideally a request is served before a new request arrives).
For the two-node topology in Fig.~\ref{fig:topology}(a), there is only one entanglement path with zero hops.
For the bottleneck topology in Fig.~\ref{fig:topology}(b), all the request initiators are on the left side of the bottleneck link, while the responders are on the right side.
As a result, the entanglement paths of the end-to-end EP have two hops (intermediate nodes).
For the AS topology in Fig.~\ref{fig:topology}(c), we create a traffic matrix such that all the entanglement paths have exactly four hops.

\subsubsection{Classical Communication Latency} 
Classical communication is needed for sending measurement results during entanglement generation, swapping, and purification, as well as for protocol coordination tasks.
It is considered a major bottleneck for quantum networks~\cite{rozped-thesis-19}.
In this paper, the classical communication latency between node $u$ and $v$ is
\begin{equation}
    l_{(u,v)} = \frac{d_{(u,v)}}{c} + \mathrm{hop}_{(u,v)} \times D_\mathrm{forward} + D_\mathrm{end~process},
    \label{eqn:latency}
\end{equation}
where $d_{(u,v)}$ is the length of the path from $u$ to $v$,
$c$ is the speed of light in optical fiber,
$\mathrm{hop}_{(u,v)}$ is the number of hops along the path,
$D_\mathrm{forward}$ is the delay for forwarding the packet at intermediate nodes, and
$D_\mathrm{end~process}$ is the delay for processing the packets at the two end nodes.
The values of $D_\mathrm{end~process}$ and $D_\mathrm{forward}$ are in the first row of Table~\ref{tab:sim-parameter}.
We assume the packet size of classical communication in quantum networks is \red{under a hundred bytes, considering the payload being a few bytes to account for a few measurement results}. 
For simplicity, we neglect transmission and queueing delays.

\subsubsection{Methods Compared}
We compare the \acp\ with an ``On Demand Only'' (\odo) strategy, where the quantum network starts generating link EPs only \emph{after} a request arrives
\red{and does not include entanglement purification.}
We also compare the \acp\ with a ``Uniform Continuous Entanglement Generation Protocol'' (\ucp), which has the same continuousness (Sec.~\ref{subsec:continuousness}) of the \acp\ but it lacks the adaptive control (Sec.~\ref{subsec:adaptive-control}), i.e. a node continuously generates EPs with neighbor nodes where the neighbors are selected by a uniform probability distribution.
Moreover, the \acp\ has two variations without entanglement purification, where multiple EPs could have been generated on the same neighbor link.
These two variations apply two different strategies: (1) select the freshest EP and (2) select a random EP.
All these strategies will be compared by two metrics: request TTS (see Eqn.~\ref{eqn:tts}) and end-to-end EP fidelity.

\begin{figure}
    \centering
    \vspace{-0.1in}
    \includegraphics[width=0.96\linewidth]{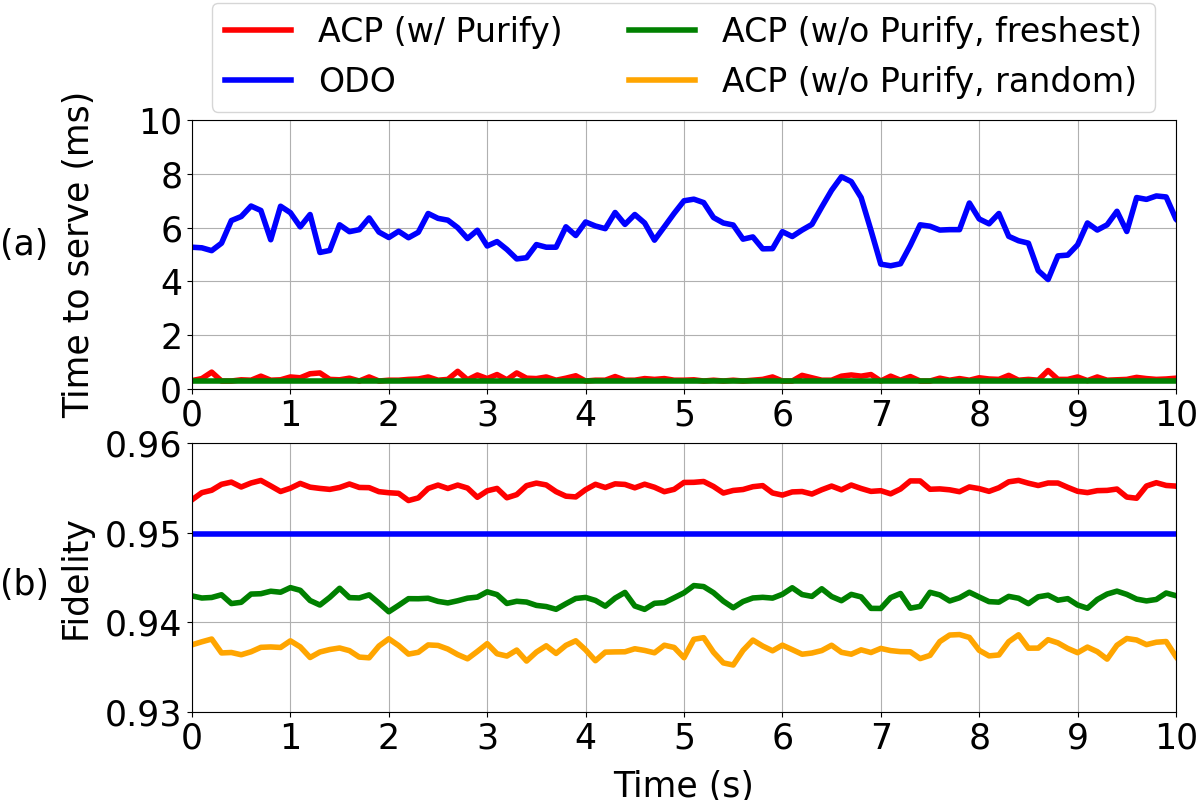}
    \vspace{-0.1in}
    \caption{Simulation results for the two-node topology network.}
    \vspace{-0.1in}
    \label{fig:2node}
\end{figure}

\subsection{Simulation Results}

\subsubsection{Two-Node Network Setting}
The \acp\ reduces the TTS by $94\%$ at most compared to the \odo.
As shown in Fig.~\ref{fig:2node}(a), the average TTS for the \odo\ is 5.95~ms, while the average TTS for the \acp\ with and without entanglement purification is 0.39~ms and 0.3~ms, respectively.
0.3~ms is the delay of a classical communication round trip time between two neighbors, see Fig.~\ref{fig:eg} and Eqn.~\ref{eqn:latency}.
The overhead of 0.09 ms is mainly because of failures in entanglement purification.
Occasionally, EPs generated by the \acp\ do not serve any request, because they are all destroyed during a failed entanglement purification attempt.
Despite a small overhead in TTS, the fidelity in Fig.~\ref{fig:2node}(b) shows an improvement of 0.01$\sim$0.02. The \acp's fidelity is 0.955, while the fidelities of the two variations of the \acp\ without entanglement purification are 0.943 and 0.937.
The \odo\ has a constant fidelity 0.949.
This simulation demonstrates that the \acp\ can not only drastically reduce the TTS, but also improve the fidelity of end-to-end EPs.

\subsubsection{20-Node Network Setting}
The \acp\ reduces the TTS by $70\%$ compared to the \odo\ and $60\%$ compared to the \ucp.
In Fig.~\ref{fig:20node}(a), the TTS for the \acp\ starts at 9~ms, then it gradually decreases to 4~ms before a change in the traffic matrix occurs. 
After the traffic matrix changes, the TTS suddenly increases to 10~ms --which is worse than the beginning-- then it gradually goes down to 4~ms again.
This trend shows the effectiveness of the adaptive control (i.e., it allows a node to dynamically adjust to changing traffic patterns).
In comparison, the TTS for the \odo\ remains oscillating around 13.3~ms, while for the \ucp\ it fluctuates at 10.3~ms.
For fidelity in Fig.~\ref{fig:20node}(b), the \acp's fidelity starts at 0.84 and it gradually increases before the traffic matrix changes, which leads to a sudden decrease in fidelity.
After this sudden decrease, the fidelity gradually increases again and reaches 0.86, which is nearly 0.02 higher than the \odo\ and the \ucp's fidelity.

\subsubsection{200-Node Network Setting}
\begin{figure}
    \centering
    \includegraphics[width=0.96\linewidth]{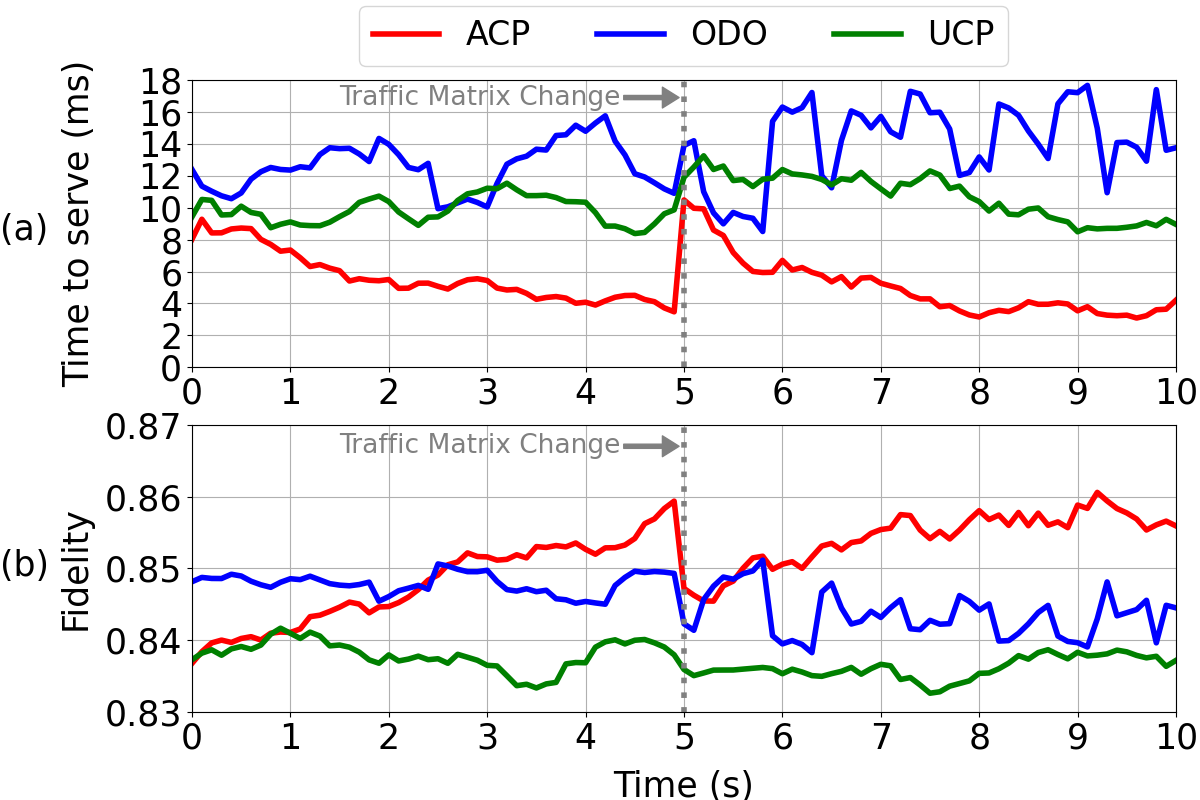}
    \vspace{-0.1in}
    \caption{Simulation results for the 20-node bottleneck topology network.}
    \vspace{-0.1in}
    \label{fig:20node}
\end{figure}

\begin{figure}
    \centering
    \includegraphics[width=0.96\linewidth]{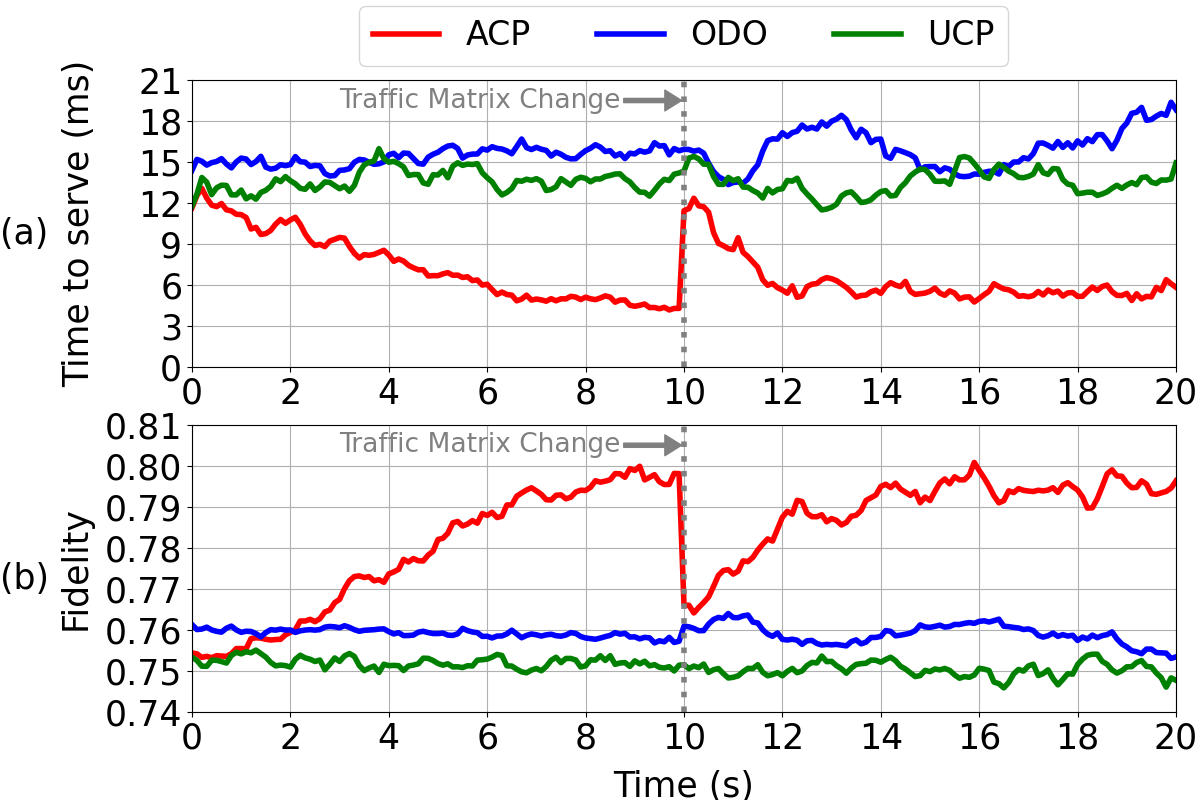}
    \vspace{-0.1in}
    \caption{\red{Simulation results for the 200-Node AS topology network.}}
    \vspace{-0.1in}
    \label{fig:200as}
\end{figure}

The trend depicted in Fig.\ref{fig:200as} is very similar to the one observed in Fig.~\ref{fig:20node}.
\red{The \acp\ (6.8~ms) reduces the TTS to $57\%$ and $50\%$ compared to the \odo\ (15.7~ms) and the \ucp\ (13.6~ms), respectively.}
Furthermore, we observe that the fidelity is increased by nearly 0.05 (see Fig.~\ref{fig:200as}(b)).
This improvement is significantly larger than the two-node and the 20-node scenarios.
This is because the fidelity improvement by the specific entanglement purification we use is the highest when the input fidelity is between 0.7 and 0.8 and in Fig.~\ref{fig:200as}(b), the fidelity of EPs happens to fall in this range.


\section{Conclusion}
\label{sec:conclusion}

In this paper, we study the Adaptive, Continuous Entanglement Generation Protocol (\acp), and implement it in the SeQUeNCe quantum network simulator. To support \acp,  
we introduce key extensions of SeQUeNCe, 
including an entanglement generation protocol that accounts for existing \acp-generated EPs, 
and an entanglement purification strategy to improve fidelity.
Through extensive simulations using SeQUeNCe, we evaluate the performance of the \acp.
Our simulation results show that the \acp\ reduces the request's time-to-serve by \red{$57\%\sim94\%$} and improves the fidelity of distributed EPs by $0.01\sim0.05$.

\bibliographystyle{IEEEtran}
\bibliography{bib}

\begin{thebibliography}{10}
\providecommand{\url}[1]{#1}
\csname url@samestyle\endcsname
\providecommand{\newblock}{\relax}
\providecommand{\bibinfo}[2]{#2}
\providecommand{\BIBentrySTDinterwordspacing}{\spaceskip=0pt\relax}
\providecommand{\BIBentryALTinterwordstretchfactor}{4}
\providecommand{\BIBentryALTinterwordspacing}{\spaceskip=\fontdimen2\font plus
\BIBentryALTinterwordstretchfactor\fontdimen3\font minus \fontdimen4\font\relax}
\providecommand{\BIBforeignlanguage}[2]{{%
\expandafter\ifx\csname l@#1\endcsname\relax
\typeout{** WARNING: IEEEtran.bst: No hyphenation pattern has been}%
\typeout{** loaded for the language `#1'. Using the pattern for}%
\typeout{** the default language instead.}%
\else
\language=\csname l@#1\endcsname
\fi
#2}}
\providecommand{\BIBdecl}{\relax}
\BIBdecl

\bibitem{wehner2018quantum}
S.~Wehner \emph{et~al.}, ``Quantum internet: A vision for the road ahead,'' \emph{Science}, vol. 362, 2018.

\bibitem{calaffi-dqc-cn24}
M.~Caleffi \emph{et~al.}, ``Distributed quantum computing: A survey,'' \emph{Computer Networks}, vol. 254, 2024.

\bibitem{zang-dqs}
A.~Zang \emph{et~al.}, ``Quantum advantage in distributed sensing with noisy quantum networks,'' \emph{arXiv preprint arXiv:2409.17089}, 2024.

\bibitem{zhan-opt-24}
C.~Zhan \emph{et~al.}, ``Optimizing initial state of detector sensors in quantum sensor networks,'' \emph{ACM Transactions on Quantum Computing}, 2024.

\bibitem{hillery-qsn-pra23}
M.~Hillery \emph{et~al.}, ``Discrete outcome quantum sensor networks,'' \emph{Phys. Rev. A}, vol. 107, 2023.

\bibitem{zhan-localization-qce23}
C.~Zhan \emph{et~al.}, ``Quantum sensor network algorithms for transmitter localization,'' in \emph{IEEE QCE}, 2023.

\bibitem{caleffi-routing-access17}
M.~Caleffi, ``Optimal routing for quantum networks,'' \emph{IEEE Access}, vol.~5, 2017.

\bibitem{shi-routing-sigcomm20}
S.~Shi \emph{et~al.}, ``Concurrent entanglement routing for quantum networks: Model and designs,'' in \emph{ACM SIGCOMM}, 2020.

\bibitem{ghader-swappingtree-tqe22}
M.~Ghaderibaneh \emph{et~al.}, ``Efficient quantum network communication using optimized entanglement swapping trees,'' \emph{IEEE TQE}, vol.~3, 2022.

\bibitem{abane-routingsurvey}
A.~Abane \emph{et~al.}, ``Entanglement routing in quantum networks: A comprehensive survey,'' 2024.

\bibitem{kolar-ac-infocom22}
A.~Kolar \emph{et~al.}, ``Adaptive, continuous entanglement generation for quantum networks,'' in \emph{IEEE INFOCOM Workshops}, 2022.

\bibitem{inesta-continuous-pra23}
A.~G. I\~nesta \emph{et~al.}, ``Performance metrics for the continuous distribution of entanglement in multiuser quantum networks,'' \emph{Phys. Rev. A}, 2023.

\bibitem{ghader-predistribution-qce22}
M.~Ghaderibaneh \emph{et~al.}, ``Pre-distribution of entanglements in quantum networks,'' in \emph{IEEE QCE}, 2022.

\bibitem{bbpssw}
C.~H. Bennett \emph{et~al.}, ``Purification of noisy entanglement and faithful teleportation via noisy channels,'' \emph{Phys. Rev. Lett.}, vol.~76, 1996.

\bibitem{sequence}
X.~Wu \emph{et~al.}, ``{SeQUeNCe}: a customizable discrete-event simulator of quantum networks,'' \emph{Quantum Science and Technology}, vol.~6, 2021.

\bibitem{chakra-routing-19}
K.~Chakraborty \emph{et~al.}, ``Distributed routing in a quantum internet,'' 2019.

\bibitem{fan-purification}
X.~Fan \emph{et~al.}, ``Distribution and purification of entanglement states in quantum networks,'' 2025.

\bibitem{pirker-multipartite-19}
A.~Pirker \emph{et~al.}, ``A quantum network stack and protocols for reliable entanglement-based networks,'' \emph{New Journal of Physics}, vol.~21, 2019.

\bibitem{fan-graphstate}
X.~Fan \emph{et~al.}, ``Optimized distribution of entanglement graph states in quantum networks,'' 2024.

\bibitem{lee-graphstate}
S.-H. Lee \emph{et~al.}, ``Graph-theoretical optimization of fusion-based graph state generation,'' \emph{Quantum}, vol.~7, 2023.

\bibitem{cicco-routing-21}
C.~Cicconetti \emph{et~al.}, ``Request scheduling in quantum networks,'' \emph{IEEE Transactions on Quantum Engineering}, vol.~2, 2021.

\bibitem{topdown-approch}
J.~F. Kurose \emph{et~al.}, \emph{Computer Networking: A Top-Down Approach (6th Edition)}, 6th~ed.\hskip 1em plus 0.5em minus 0.4em\relax Pearson, 2012.

\bibitem{quisp}
R.~Satoh \emph{et~al.}, ``{QuISP}: a quantum internet simulation package,'' in \emph{IEEE QCE}, 2022.

\bibitem{netsquid}
T.~Coopmans \emph{et~al.}, ``{NetSquid}, a network simulator for quantum information using discrete events,'' \emph{Communications Physics}, 2021.

\bibitem{zang2022simulation}
A.~Zang \emph{et~al.}, ``Simulation of entanglement generation between absorptive quantum memories,'' in \emph{IEEE QCE}, 2022.

\bibitem{davossa2024simulation}
L.~d'Avossa \emph{et~al.}, ``Simulation of quantum transduction strategies for quantum networks,'' 2024.

\bibitem{wu2024parallel}
X.~Wu \emph{et~al.}, ``Parallel simulation of quantum networks with distributed quantum state management,'' \emph{ACM TOMACS}, vol.~34, 2024.

\bibitem{matsuo-ruleset-pra19}
T.~Matsuo \emph{et~al.}, ``Quantum link bootstrapping using a ruleset-based communication protocol,'' \emph{Phys. Rev. A}, vol. 100, 2019.

\bibitem{ent_dist_epp}
A.~Zang \emph{et~al.}, ``Entanglement distribution in quantum repeater with purification and optimized buffer time,'' in \emph{INFOCOM Workshops}, 2023.

\bibitem{dur2003entanglement}
W.~D{\"u}r \emph{et~al.}, ``Entanglement purification for quantum computation,'' \emph{Physical Review Letters}, vol.~90, no.~6, p. 067901, 2003.

\bibitem{zang2024no}
A.~Zang \emph{et~al.}, ``No-go theorems for universal entanglement purification,'' \emph{arXiv preprint arXiv:2407.21760}, 2024.

\bibitem{rozped-thesis-19}
F.~Rozpedek, ``Building blocks of quantum repeater networks,'' Ph.D. dissertation, Delft University of Technology, 2019.

\end{thebibliography}

\end{document}